\documentstyle[preprint,epsfig,prb,aps]{revtex}

\begin{document}

\title{\bf The influence of the boundary resistivity on the proximity effect}

\author{C. Ciuhu and A. Lodder}

\address{Faculty of Sciences / Natuurkunde en Sterrenkunde, 
Vrije Universiteit,
         De Boelelaan 1081,\\ 1081 HV Amsterdam, The Netherlands}

\date{\today}
\maketitle
\begin{abstract}
{
We apply the theory of Takahashi and Tachiki in order to explain 
theoretically the dependence of the upper critical magnetic field 
of a S/N multilayer on the temperature. 
This problem has been already investigated in the literature, 
but with a use of 
an unphysical scaling parameter for the coherence length. 
We show explicitely that, in order to describe the data, 
such an unphysical parameter is unnecessary 
if one takes into account the boundary resisitivity of the S/N interface. 
We obtain a very good agreement with the experiments 
for the multilayer systems Nb/Cu and V/Ag, with various layer thicknesses. 
}
\end{abstract}

\section{Introduction}
\label{intro}

In trying to describe the experimental 
data for different kinds of multilayers, such as Nb/Cu or V/Ag, 
Koperdraad calculated upper critical magnetic fields versus temperature, 
using the Takahashi-Tachiki theory for infinite multilayers \cite{taka}. 
He used as fitting parameters the bulk critical temperature of 
the S-layer $T_c^S$, the ratio between the 
densities of states of the two materials $N_S/N_N$, and the 
two corresponding diffusion 
constants, $D_S$ and $D_N$. 

In calculating the magnetic field anisotropy, which is the ratio between the 
parallel and perpendicular upper critical magnetic fields 
$H_{c2,\parallel}/H_{c2,\perp}$, two choices 
were possible for the diffusion constants, 
which lead to two solutions, 
called the first and the 
second solution \cite{adri,rutg,kopwulod,rutger2,lodkop,koperN}. 
For the first solution, the fitted parameters are close to what one 
knows from the measurements. 
However, the dimensional crossover, typical for S/N multilayers,  
appeared to lie at a much higher temperature than the measured one. 
In the second solution the upper parallel critical 
magnetic field exhibits a dimensional crossover at a lower 
temperature than the experimental one. 
A characteristic of this type of solution is that the superconductivity 
nucleation point for the parallel magnetic field shifts from the S-layer at 
low temperatures to the N-layer at higher temperatures, which seems 
unphysical for a S/N multilayer whose $T_c^N=0$. 
Another unphysical aspect is 
that the fitted critical temperature for the S-layer 
is larger than the one known for the bulk (8.9 K). 
Moreover, instead of an expected concave 
2D-aspect of the curve at lower temperatures, the 
calculations lead to a convex type of curve. 
In order to fit the experimentally observed dimensional crossover 
with the theoretical one, Koperdraad introduced 
a scaling parameter $\alpha$ for the magnetic coherence length. 
However, the physical interpretation for this free parameter 
remains an open question. 

Looking for a physical factor which can replace the role of the unphysical 
scaling parameter in fitting the data, 
Aarts \cite{aarts} suggested to consider finite 
samples rather than the infinite ones 
on which Koperdraad did his calculations. 
In finite samples one has to face surface effects. 
Model calculations done on finite samples\cite{calina} show that 
the surface nucleation of the superconductivity is more pronounced for 
multilayers with thinner layers, but it almost disappears as one increases 
the thickness of the layers. 
Since the fitting problem mentioned above showed up particularly 
for thick-layer systems, 
taking into account surface superconductivity does not bring any 
essential improvement to the already existing results. 

In the present paper we consider the influence of an S/N interface resistivity, 
in order to get rid of the unphysical parameter $\alpha$. 
This is in line with experimental evidence that the 
interfaces of artificial multilayers for metals with a different 
crystal structure such as Nb/Cu are quite rough \cite{schuller}. 
Indeed, we find that a finite boundary 
resistivity ($R_B$) allows for a good fit with the experimental data. 

The paper is organized as follows. In Section \ref{boundres} we summarize 
the theory of Takahashi and Tachiki, and we introduce the boundary 
resistivity by adjusting the boundary conditions. We also illustrate 
the role of the boundary resistivity on the proximity effect. Section 
\ref{results} is dedicated to the numerical results and conclusion. 

\section{Theory including boundary resistivity}
\label{boundres}

First we summarize the Takahashi-Tachiki theory for S/N multilayers. 
The theory starts from the Gor'kov 
equation \cite{gork} 
for the pair potential $\Delta( {\bf r})$, 
with a space-dependent coupling constant $V({\bf r})$, 
\begin{equation}
\label{gorkov}
\Delta ({\bf r})=V( {\bf r})kT \sum_{ \omega} \int d^{3} {\bf r'}
Q_{ \omega}( {\bf r,r'})\Delta ( {\bf r'}),
\end{equation}
in which the summation runs over the Matsubara frequencies. 
By averaging over the impurity configurations and considering the dirty 
limit, it was shown that the integration kernel $Q_{\omega}$ obeys
a Green's function like equation 
\begin{equation}
\label{green}
[2|\omega |+L({\bf \nabla})]Q_{\omega }({\bf r,r'})=2\pi N({\bf r})\delta
({\bf r-r'}),
\end{equation}
where 
\begin{equation}
\label{operator}
L({\bf \nabla })=- \hbar D({\bf r})({\bf \nabla} -\frac {2ie}{\hbar c }{\bf A}
({\bf r}))^2.
\end{equation}
This result appears to be equivalent with a different approach going 
back to Usadel \cite{koperN,usadel}. 
The material parameters $V({\bf r})$, $N({\bf r})$, and $D({\bf r})$ are
the BCS coupling constant, 
the electronic density of states at the Fermi energy
and the diffusion coefficient, respectively. 
In practice, they are treated as being a constant in each single layer. 
At the interfaces de Gennes boundary conditions are 
imposed,\cite{degen} which require the continuity of
$\frac{F({\bf r})}{N({\bf r})}$ and $D({\bf r})({\bf \nabla} -\frac {2ie}
{\hbar c }{\bf A}({\bf r}))F({\bf r})$, where the pair amplitude
$F({\bf r})$ is related to the gap function $\Delta ({\bf r})$ through
\begin{equation}
\label{pair-gap}
\Delta ({\bf r})=V({\bf r})F({\bf r}).
\end{equation}

Takahashi and Tachiki provide a way of solving Eqs.
(\ref{gorkov}) and (\ref{green}) 
by developing the kernel $Q_{\omega }({\bf r,r'})$ and the pair function
$F({\bf r})$ in terms of a complete set of eigenfunctions of the 
differential operator $L({\bf \nabla })$. 
These eigenfunctions are labeled by the parameter
$\lambda$, and the eigenvalues are $\epsilon_{\lambda }$. They are solution 
of the eigenvalue problem
\begin{equation}
\label{lambda}
L(\nabla)\Psi_{\lambda }= \epsilon_{\lambda }\Psi_{\lambda }, 
\end{equation}
subject to de Gennes boundary conditions. 
The requirement of the existence of a solution for Eq. (\ref{gorkov})
leads to the equation
\begin{equation}
\label{det}
\det|\delta _{\lambda \lambda '}-2\pi kT\sum _{\omega }\frac{1}{2|\omega |+
\epsilon_{\lambda }}V_{\lambda \lambda '}|=0.
\end{equation}

For finite multilayers in vacuum, the de Gennes boundary conditions 
insure that there is no current flow through the interface 
between the multilayer and the vacuum. These boundary conditions read 
$D({\bf r})({\bf \nabla} -\frac {2ie}
{\hbar c }{\bf A}({\bf r}))|_zF({\bf r})=0$ for the pair amplitude, at the 
interface with the vacuum. 
As usual for these type of layered systems, the growth direction
coincides with the $z$ direction. 
When applied to the eigenfunctions $\Psi_{\lambda }$, they become 
$\frac{\partial \Psi _{\lambda}(x,y,z)}{\partial z}=0$, where we made use of 
the gauge ${\bf A}({\bf r})=(Hz,0,0)$ when the magnetic field is applied 
parallel to the layers, and ${\bf A}({\bf r})=(0,Hx,0)$ for the perpendicular 
magnetic field. 
In the absence of a magnetic field, the solution of Eq. (\ref{det}) giving the
largest value for the critical temperature is the physical one.
In the presence of a field, solving this equation allows us to derive
the $H_{c2} (T)$ curves.
The temperature at which $H_{c2}\rightarrow 0$ is $T_c$.

A further step in applying the theory of Takahashi and Tachiki is to 
consider the effect of the S/N interface resistivity. 
In our calculations, we make use of more general boundary 
conditions rather than the de Gennes ones. 
Such boundary conditions were investigated by Kupriyanov and Lukichev 
\cite{kupriyanov} and according to Golubov \cite{GolKup} and 
Khushainov \cite{Khushainov} they can be written as: 

\begin{equation}
\label{boundary}
D({\bf r})\frac{\partial}{\partial z}F({\bf r})|_{\bf r=r^+}=
D({\bf r})\frac{\partial}{\partial z}F({\bf r})|_{\bf r=r^-}=
\frac{1}{e^2R_B}(\frac{F({\bf r^+})}{N({\bf r^+})}-\frac{F({\bf r^-})}
{N({\bf r^-})}). 
\end{equation}
The boundary resistivity $R_B$ is a parameter which characterizes the 
barrier which electrons encounter at the interface. A source of this 
resistance comes from the mismatch of the Fermi (or electronic) 
levels, lattice structure and lattice constant of the two composite 
metals. As a consequence, $R_B$ reduces the migration 
of the Cooper pairs from the S-layer to the N-layer, 
by that diminishing the proximity effect. 

As we will illustrate in the following, $R_B$ modifies 
the critical temperature $T_c$ of the multilayer and the 
magnetic field anisotropy, defined as $H_{c2,\parallel}/H_{c2,\perp}$. 
By consequence, including $R_B$ as a parameter, the two solutions 
used by Koperdraad have to be reconsidered. 
It will turn out that in using the boundary resistivity as a free 
parameter, only one solution will be possible for the fitting, 
instead of the two solutions. 
This solution fits the experimental data, without using any other free 
parameter, such as the scaling-parameter $\alpha$. 

Let us first consider the situation in which there is no magnetic 
field applied to the system. 
As mentioned already, a finite 
boundary resistivity reduces 
the proximity effect. 
This leads to a higher multilayer critical temperature than in 
the case of perfect transparency of the interfaces. 
As a consequence, the bulk critical temperature $T_c^S$ used to fit 
the multilayer critical temperature will be smaller than 
the one used by 
Koperdraad. This leads us in a good direction, since the previously 
used $T_c^S$ was higher than the measured value. 

As an illustration of the influence of $R_B$ on the proximity effect, 
we calculate  the dependence of the critical temperature of 
a multilayer on the thickness of the layers for different choices 
for the boundary resistivity. The results 
for an 11-layers Nb/Cu system are shown in FIG. \ref{prox1}. 
First, one notices that 
as the layer thickness decreases, the multilayer critical temperature 
converges smoothly towards 0, whereas in the thick layer limit, 
it converges to the bulk critical temperature $T_c^S$. 
Further, the curves show that below a certain thickness of the layers, 
$d_{cr}$, the superconductivity disappears. Moreover, this critical 
thickness $d_{cr}$ decreases with the increasing of the boundary resistivity, 
illustrating the fact that due to $R_B$, the density of Cooper pairs 
is more localized in the 
S-layers of the multilayer, so that the system becomes a better 
superconductor. 

We consider now the presence of a magnetic field. 
When a magnetic field is applied to the system perpendicularly to the 
interfaces, due to the in-plane motion of the Cooper pairs, the influence 
of the boundary resistivity is weak. However, for the magnetic field 
parallel to the interfaces, the picture looks different. In this situation, 
the Cooper pairs move such that they cross the interface, which means that 
they experience the influence of the boundary resistivity much 
stronger. 
In the presence of a boundary 
resistivity, the diffusion of the Cooper pairs from the 
S-layers into the N-layers is diminished. 
The proximity effect is weaker, leading to a 
higher critical temperature for the same magnitude of the magnetic field. 
Thus we can conclude that the boundary resistivity 
increases the anisotropy ratio $H_{c2,\parallel}/H_{c2,\perp}$. 

In addition it appears 
that the dimensional crossover temperature is shifted towards 
higher temperatures. This means that 
the first solution is not favourable, whereas the second solution has chances 
to be ameliorated.

In the following section we will take as a starting point the second 
solution and we will present the corrections which are performed in view of 
a fitting with the experimental data. 

\section{Results and Conclusions}
\label{results}

Considering the second solution, its inconveniences consist 
in the fact that at low temperature the $H_{c2,\parallel}(T_c)$ 
curve is convex, 
instead of the well known concave square-root behaviour for the 2D systems. 
Besides, at high temperatures the nucleation of the superconductivity 
lies in the N-layer, which is unphysical for such S/N systems. 
Moreover, a too large ratio 
$N_ND_N/N_SD_S$ is used in fitting, in order to obtain the 
corresponding anisotropy. 

All these shortcomings are remedied by considering 
a finite boundary resistivity. 
In FIG. \ref{ParaNbCu1} we show results for Nb(171\AA)/Cu(376\AA) multilayer. 
The solid curves are obtained by accounting for a finite $R_B$. 
The dashed curves are Koperdraad's result, which could be improved by using 
a scaling parameter, still lacking a physical interpretation. 
The perpendicular field curves are not very sensitive to the change 
of the parameters. 
We fitted the points $H_{c2}(T_{c2})=0$, 
$H_{c2,\parallel}(T^{DCO})$, and $H_{c2,\perp}(T^*)$ 
on the measured critical field curves, 
rather than the points $H_{c2}(T_{c2})=0$, $H_{c2,\parallel}(T^*)$, 
and $H_{c2,\perp}(T^*)$, used by Koperdraad. 
Here $T^{DCO}$ is the temperature where the dimensional 
crossover occurs on the parallel magnetic field curve, 
and $T^*$ corresponds to the experimental point at the lowest temperature. 
In Tabel \ref{tabel}, we show the data 
used in our fitting ($T_c^S$, $D_S$, $D_N$, and $R_B$), 
compared to the data used by Koperdraad 
($T_c^{S,K}$, $D_S^K$, and $D_N^K$). 
For example, in fitting the Nb(171\AA)/Cu(376\AA) system, 
we used $D_S=2.4$ cm$^2$/s, $D_N=78$ cm$^2$/s and 
$R_B=3.17\mu\Omega$cm, instead of 
$D_S=0.65$ cm$^2$/s and $D_N=138$ cm$^2$/s, used by Koperdraad {\it et al.} 
The latter set is rather unrealistic, while the first set compares nicely 
with the diffusion constants used by Biagi {\it et al. }\cite{biagi} 
The resistivity has the same order of magnitude as the resistivity of Nb 
at 77K, which is $\rho_{Nb}=3 \mu\Omega$cm, 
and it is an order of magnitude larger than the Cu value of 
0.2 $\mu\Omega$cm. 
Since the interface can be considered as a dirty mixture, 
the value of $R_B$ looks reasonable. 
The use of a smaller and more realistic ratio $N_ND_N/N_SD_S$, can be 
explained as follows. In the absence of a boundary resistivity, $R_B=0$, 
the anisotropy at a certain temperature $T^*$ is directly related 
to the ratio $N_ND_N/N_SD_S$. However, the anisotropy increases 
when one considers a finite $R_B$, so that a smaller ratio $N_ND_N/N_SD_S$ 
is necessary to fit the anisotropy of the upper critical fields. 
Besides, as one can notice in FIG. \ref{ParaNbCu1}, 
this choice of the diffusion constants is such that 
the convex behaviour of Koperdraad's $H_{c2,\parallel}(T)$ curve is 
turned into a concave one, characteristic for a 2D system. 
Furthermore, in our solution 
the nucleation of the superconductivity takes place in the 
S-layer, as one expects from physical reasons. 
Clearly, a good agreement between theory and measurements is obtained. 

In the same way we fitted the data for two other Nb/Cu 
multilayers, as well as for a V/Ag system. 
The results are shown in FIG. \ref{ParaNbCu2} for Nb(172\AA)/Cu(333\AA), 
in FIG. \ref{ParaNbCu3} for Nb(168\AA)/Cu(147\AA), and in 
FIG. \ref{ParaVAg} for V(240\AA)/Ag(480\AA) multilayer. 
The experimental data are taken from the literature \cite{chun,kanoda}. 

In conclusion, by focusing on a fit at the dimensional crossover 
temperature and allowing for a finite boundary resistivity, the theory 
describes the experimental data nicely. By that the merit of the 
scaling parameter introduced by Koperdraad {\it et al.} is reduced 
considerably, the more so as up to now this parameter was not assigned 
with any physical interpretation. A finite boundary resistivity, on the 
other hand, is in accordance with experimental evidence \cite{schuller}. 

\begin{table}
\begin{tabular}{l|c|c|c|c|c|c|c}
The system & $T_c$[K] & $D_S$[cm$^2$/s] & $D_N$[cm$^2$/s] & $R_B$
[$\mu\Omega$cm] & $T_c^K$[K] & $D_S^K$[cm$^2$/s] & $D_N^K$[cm$^2$/s] \\
\hline  
Nb(171\AA)/Cu(376\AA)  & 9.20 & 2.4  &  78 & 3.17 & 9.89 & 0.65 &  138 \\
Nb(172\AA)/Cu(333\AA)  & 9.20 & 1.23 &  69 & 2.07 & 9.88 & 0.64 &  180 \\
Nb(168\AA)/Cu(147\AA)  & 9.50 & 1.45 &  73 & 2.38 & 9.61 & 0.58 &  231 \\
V(240\AA)/Ag(480\AA)   & 5.47 & 1.1  &  54 & 3.52 & 5.70 & 0.67 & 73.4   
\end{tabular}
\label{tabel}
\end{table}

\section*{acknowledgement}

One of the authors (CC) would like to thank Dr. J. Aarts for 
useful discussions.

\newpage

\section*{CAPTIONS AND FIGURES}

\begin{figure}[htb]
\centerline{\epsfig{figure=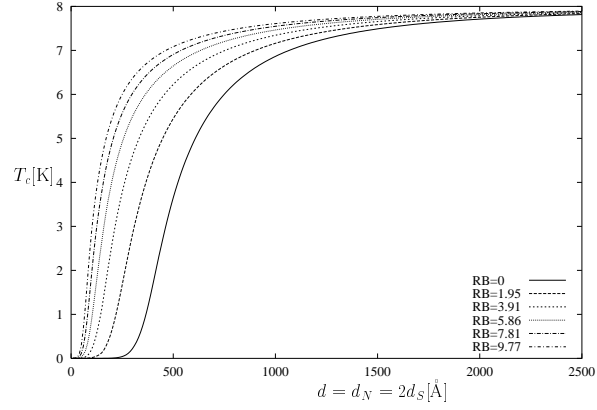,angle=90,width=8.0cm}}
\caption[]{The critical temperature $T_c$ 
for an 11-layers Nb/Cu system, as a function of the 
layer thicknesses for different values of the boundary resistivity $R_B$, 
measured in $\mu\Omega$cm. }
\label{prox1}
\end{figure}

FIG. \ref{prox1}. 
The critical temperature $T_c$ 
for an 11-layers Nb/Cu system, 
as a function of the layer thickness, 
for different values of the boundary resistivity $R_B$, 
measured in $\mu\Omega$cm. 

\begin{figure}[htb]
\centerline{\epsfig{figure=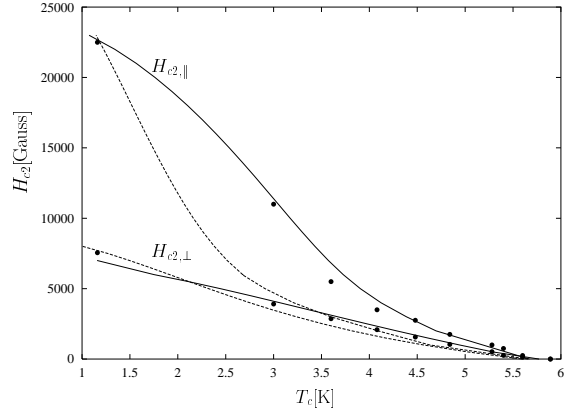,angle=90,width=8.0cm}}
\caption[]{
The upper parallel and perpendicular magnetic fields 
for the multilayer Nb(171\AA)/Cu(376\AA). The dots denote the 
experimental points \cite{chun}. }
\label{ParaNbCu1}
\end{figure}

FIG. \ref{ParaNbCu1}. 
The upper parallel and perpendicular magnetic fields 
for the multilayer Nb(171\AA)/Cu(376\AA). 
The dots denote the experimental points \cite{chun}. 

\begin{figure}[htb]
\centerline{\epsfig{figure=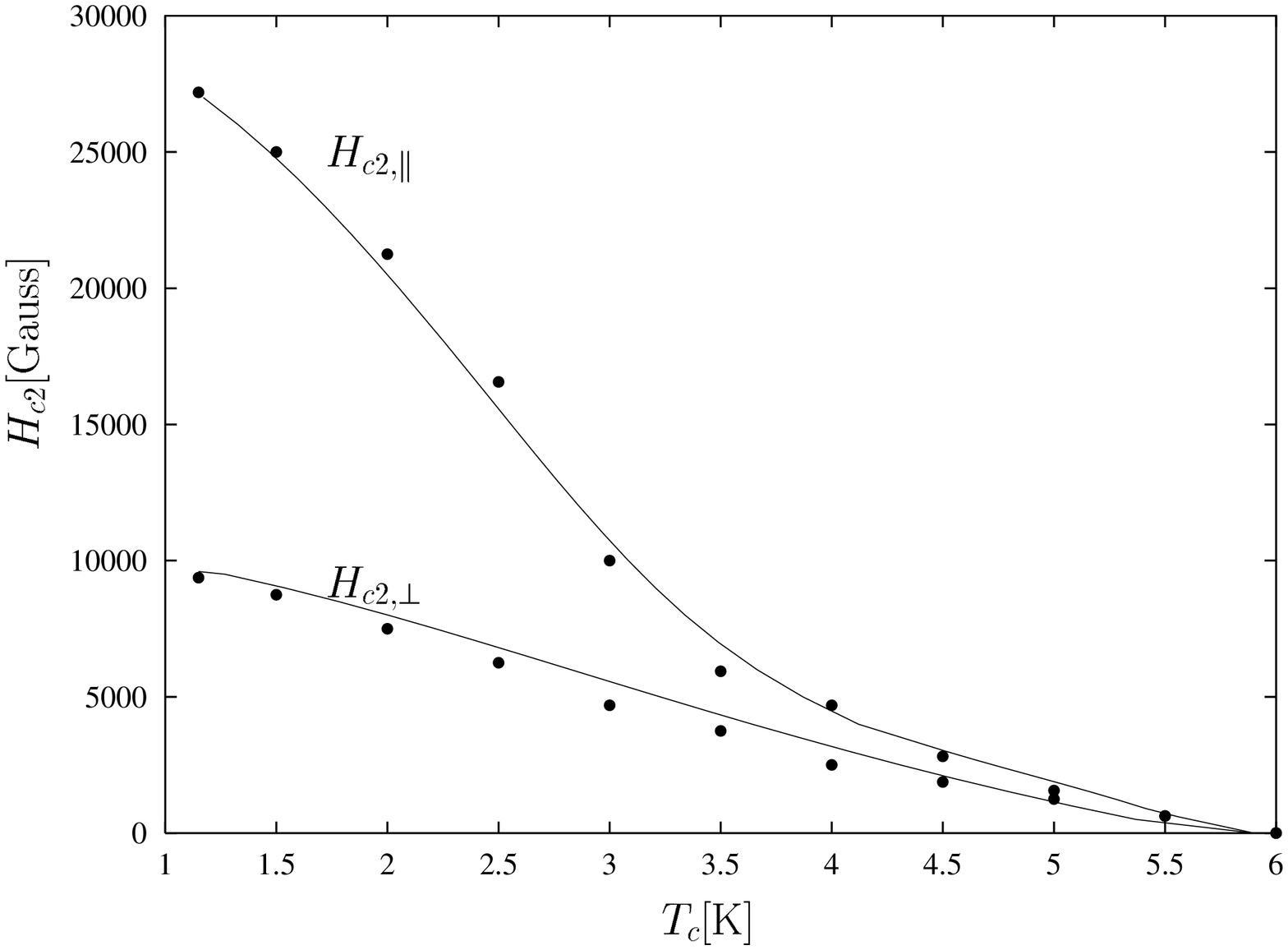,angle=90,width=8.0cm}}
\caption[]{
The upper parallel and perpendicular magnetic fields experimental \cite{chun} 
and theoretical curves 
for the multilayer Nb(172\AA)/Cu(333\AA).} 
\label{ParaNbCu2}
\end{figure}

FIG. \ref{ParaNbCu2}. 
The upper parallel and perpendicular magnetic fields experimental \cite{chun} 
and theoretical curves 
for the multilayer Nb(172\AA)/Cu(333\AA). 

\begin{figure}[htb]
\centerline{\epsfig{figure=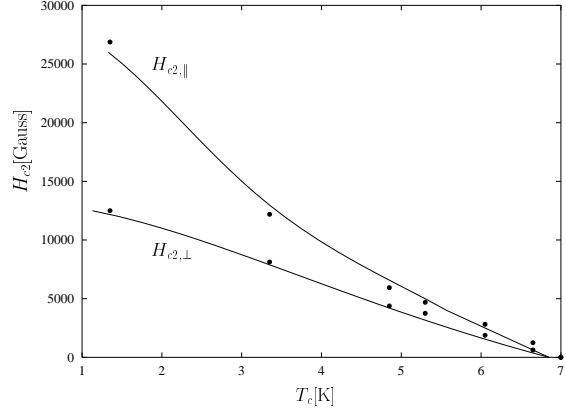,angle=90,width=8.0cm}}
\caption[]{
The upper parallel and perpendicular magnetic fields experimental \cite{chun} 
and theoretical curves 
for the multilayer Nb(168\AA)/Cu(147\AA).} 
\label{ParaNbCu3}
\end{figure}

FIG. \ref{ParaNbCu3}. 
The upper parallel and perpendicular magnetic fields experimental \cite{chun} 
and theoretical curves 
for the multilayer Nb(168\AA)/Cu(147\AA). 

\begin{figure}[htb]
\centerline{\epsfig{figure=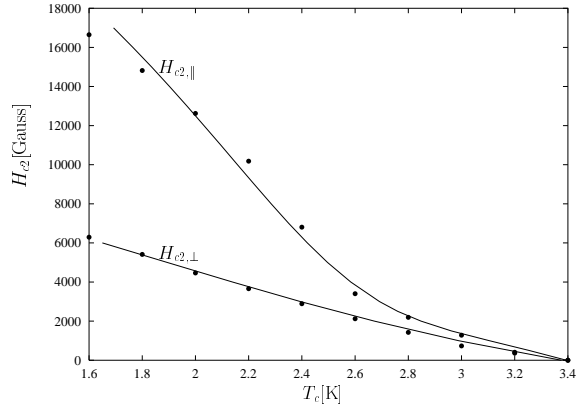,angle=90,width=8.0cm}}
\caption[]{
The upper parallel and perpendicular magnetic fields experimental 
\cite{kanoda} 
and theoretical curves for the multilayer V(240\AA)/Ag(480\AA).} 
\label{ParaVAg}
\end{figure}

FIG. \ref{ParaVAg}. 
The upper parallel and perpendicular magnetic fields experimental \cite{kanoda} 
and theoretical curves 
for the multilayer V(240\AA)/Ag(480\AA). 

\newpage

\end{document}